\documentclass[twocolumn,showpacs,amsmath,amssymb]{revtex4}
\usepackage{graphicx}
\newcommand{\ltsim}{\protect\raisebox{-0.5ex}{$\:\stackrel{\textstyle <}{\sim}\:$}}
\newcommand{\gtsim}{\protect\raisebox{-0.5ex}{$\:\stackrel{\textstyle >}{\sim}\:$}}
\begin{document}

\title{Enhancement of phonon effects in photoexcited states of one-dimensional Mott insulators}
\author{H. Matsueda${}^{a}$}
\author{A. Ando${}^{b}$}
\author{T. Tohyama${}^{c}$}
\author{S. Maekawa${}^{b,d}$}
\affiliation{
${}^{a}$Sendai National College of Technology, Sendai 989-3128, Japan \\
${}^{b}$Institute for Materials Research, Tohoku University, Sendai 980-8578, Japan \\
${}^{c}$Yukawa Institute for Theoretical Physics, Kyoto University, Kyoto 606-8502, Japan \\
${}^{d}$CREST, Japan Science and Technology Agency (JST), Kawaguchi 332-0012, Japan}
\date{\today}
\begin{abstract}
We examine how the electron correlation affects the electron-phonon (EP) interaction in the linear optical absorption spectrum of the one-dimensional (1D) extended Hubbard-Holstein model. A density matrix renormalization group (DMRG) calculation shows that the effect of the EP interaction on an exciton is enhanced by increasing the on-site Coulomb repulsion. This enhancement is in contrast to the effect of the EP interaction on the ground state where the Peierls instability is suppressed by the on-site Coulomb repulsion. The DMRG data with the EP interaction fit with absorption experiments in 1D cuprates better than those for the extended Hubbard model.
\end{abstract}
\maketitle

The interplay of strong electron-electron correlation and electron-phonon (EP) interaction has attracted much attention~\cite{Iadonisi}. One of the issues is to understand the effect of the EP interaction on carriers doped into a Mott insulator. In the ground state of the Mott insulator, the on-site Coulomb repulsion makes the charge distribution uniform, and then the EP interaction is irrelevant. However, the mobile carriers created in the Mott insulator by chemical and/or photo dopings induce the charge fluctuation, and thus the EP interaction is expected to play an important role in various physical quantities~\cite{Bauml,Capone,Cappelluti,Tsutsui,Fehske,Ning,Matsueda3,Cappelluti2}.

In this paper, we address this issue in the photoexcited states of the one-dimensional (1D) Mott insulators. The effect of the EP interaction on the photocarriers can be observed by optical measurements. In the Raman spectroscopy for Ca${}_{1.8}$Sr${}_{0.2}$CuO${}_{3}$, Raman-forbidden phonon peaks are resonantly enhanced for charge-transfer excitation~\cite{Yoshida}. This observation may indicate the change of the selection rule by coupling between phonons and photocarriers created by the charge-transfer excitation. One of the phonon modes has the local symmetry equal to a breathing mode whose effect on the single-particle excitation spectra has been intensively studied in high-$T_{c}$ cuprates~\cite{Mishchenko,Rosch}.

In the photoexcited state of the Mott insulator, there are two types of photocarriers called 'holon' and 'doublon' representing empty and doubly occupied sites, respectively~\cite{Maekawa}. Usually, a holon and a doublon form an exciton due to the intersite Coulomb repulsion~\cite{Stephan,Matsueda}. However, this exciton is different from an electron-hole pair in semiconductors, since the strong electron correlation prohibits the exchange of holon and doublon. It has been pointed out that this prohibition enhances a magnitude of the third-order optical nonlinear susceptibility~\cite{Mizuno,Kishida2}. Thus, it is important to understand how the exciton in the Mott insulators is modified by the EP interaction.

We examine the effect of the EP interaction on the exciton in the 1D extended Hubbard-Holstein model. We perform the density matrix renormalization group (DMRG) calculation of the optical spectra. The DMRG calculation shows that the effect of the EP interaction on the exciton is enhanced by increasing the on-site Coulomb repulsion. This enhancement is in contrast to the effect of the EP interaction on the ground state where the Peierls instability is suppressed by the on-site Coulomb repulsion. The DMRG data with the EP interaction fit with absorption experiments in 1D cuprates better than those for the extended Hubbard model~\cite{Ono}.

The coupling between an electron and a breathing phonon in the cuprates can be mapped onto the Holstein- and Peierls-type interactions~\cite{Gunnarsson}. Since the Holstein interaction is stronger than the Peierls one, we start with the 1D Hubbard-Holstein model. In addition, we introduce the nearest neighbor Coulomb repulsion which leads to the formation of an exciton. The Hailtonian is defined by
\begin{eqnarray}
H &=& -t\sum_{i,\sigma}(c_{i,\sigma}^{\dagger}c_{i+1,\sigma}+{\rm H.c.}) \nonumber \\
&& + U\sum_{i}n_{i,\uparrow}n_{i,\downarrow} + V\sum_{i}(n_{i}-1)(n_{i+1}-1) \nonumber \\
&& + \omega_{0}\sum_{i}b_{i}^{\dagger}b_{i} - g\sum_{i}(b_{i}^{\dagger}+b_{i})(n_{i}-1),
\label{H}
\end{eqnarray}
where $c_{i,\sigma}^{\dagger}$ ($c_{i,\sigma}$) is a creation (annihilation) operator of an electron at site $i$ with spin $\sigma$, and $b_{i}^{\dagger}$ ($b_{i}$) is a creation (annihilation) operator of a phonon at site $i$. This model includes electron hopping, $t$, on-site and nearest-neighbor Coulomb repulsions, $U$ and $V$, respectively, phonon frequency, $\omega_{0}$, and EP coupling, $g$. Here, we neglect the dispersion of the phonon in order to make our discussion simple.

We calculate the current-current correlation function defined by
\begin{eqnarray}
\chi(\omega)=-\frac{1}{\pi L}{\rm Im}\left<0\left|j\frac{1}{\omega+E_{0}-H+i\gamma}j^{\dagger}\right| 0\right>,
\end{eqnarray}
where $\left|0\right>$ is the ground state with energy $E_{0}$, $L$ is the system size, the current operator $j$ is defined by $j=-it\sum_{i,\sigma}(c_{i,\sigma}^{\dagger}c_{i+1,\sigma}-{\rm H.c.})$, and $\gamma$ is a small positive number. The optical absorption spectrum is proportional to $\chi(\omega)/\omega^{2}$.

In order to calculate $\chi(\omega)$, the infinite-system DMRG method~\cite{White} is applied to an open boundary system. We have also carried out the finite-system DMRG calculation for some data, and this calculation quantitatively agrees with the infinite-system DMRG data. In eq. (\ref{H}), we have introduced the $V$-term, $V(n_{i}-1)(n_{i+1}-1)$, instead of the standard notation, $Vn_{i}n_{i+1}$, and the present notation does not produce the boundary effect which is localization of holon and doublon at the edge. We introduce a mixed-state density matrix which is composed of the ground state $\left|0\right>$, a photoexcited state $j^{\dagger}\left|0\right>$, and two correction vectors $(\epsilon+E_{0}-H+i\gamma)^{-1}j^{\dagger}\left|0\right>$ with $\epsilon=\omega, \omega+2\gamma$. In terms of the reduced Hilbert space made by the density matrix, the spectrum for $\omega<\epsilon<\omega+2\gamma$ is calculated after one DMRG run. The number of the eigenstates of this density matrix are taken up to 300 for renormalization. In order to correctly deal with the phononic degrees of freedom, we use the pseudo-boson method~\cite{Jeckelmann2}. In this method, a boson operator with a restricted Hilbert space is exactly transformed into a set of hard-core bosons. For instance, the boson operator with 4 states is transformed into $b^{\dagger}=a_{1}^{\dagger}+\sqrt{2}a_{2}^{\dagger}a_{1}+(\sqrt{3}-1)a_{1}^{\dagger}a_{2}^{\dagger}a_{2}$ with use of the hard-core boson operators $a_{1}$ and $a_{2}$. Then, step-by-step renormalization of each boson is carried out. For $g\le\sqrt{0.5}t$, we have taken 4 hard-core bosons per site, which correspond to 16 phonon states per site.

In numerical calculations, we take $L=20$, $20$ electrons (half-filling), $U=10t$, $\omega_{0}=0.5t$, and $\gamma=0.2t<\omega_{0}$. The magnitude of $\omega_{0}$ may be three-times larger than the frequency of the breathing mode in cuprates~\cite{Gunnarsson} in order to make numerical calculation feasible. However, in the Hubbard-Holstein model, a dimensionless parameter, $\lambda=g^{2}/4t\omega_{0}$, is more important for estimation of the strength of the EP interaction, and thus $\lambda$ is used for comparison of numerical data with observed optical spectra.

\begin{figure}[htbp]
\begin{center}
\includegraphics[width=11cm]{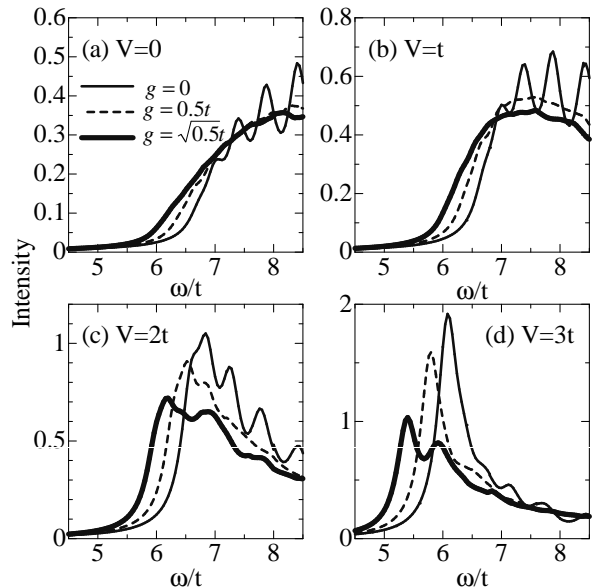}
\end{center}
\caption{Current-current correlation function $\chi(\omega)$ for the Hubbard-Holstein model with $U=10t$. (a) $V/t=0$, (b) $V/t=1$, (c) $V/t=2$, and (d) $V/t=3$.}
\label{fig1}
\end{figure}

In Fig.~\ref{fig1}, we show $\chi(\omega)$ for various $g$ and $V$ values. For $g=0$, the $V$ dependence of the spectra has been studied intensively, and the present data are consistent with the previous data~\cite{Matsueda}: for $V<2t$ only a continuum band exists, which is composed of nearly free holon and doublon, while for $V\ge 2t$ the holon and doublon form an excitonic bound state. A pronounced peak at $\omega=6.1t$ in Fig. 1(d) and a tiny hump at $\omega=6.6t$ in Fig. 1(c) correspond to the exciton (the other spiky structures in Figs. 1(a)-(d) come from the finite-size effect). For $g\ne 0$, the DMRG data show that the effect of $g$ on the spectra becomes remarkable with increasing $V$, and in particular the bound exciton for $V\ge 2t$ is strongly renormalized by the EP interaction.

For $V<2t$ and $g\ne 0$, the band edge shifts toward lower-energy region. This is because the holon and doublon are dressed with phonons, leading to an energy gain of the photoexcited state. The energy gain estimated by the DMRG data is smaller than $2g^{2}/\omega_{0}$ given by the second-order perturbation expansion with respect to $g$ for $t=0$. For instance, the energy gain for $V=t$ and $g=\sqrt{0.5}t$ in Fig. 1(b) is about $0.5t$, and this is smaller than $2g^{2}/\omega_{0}=2t$. In this case, the motion of holon and doublon suppresses the effect of $g$ on these carriers.

For $V\ge 2t$ and $g\ne 0$, the exciton peak for $g=0$ splits into multiple peaks. For $g=\sqrt{0.5}t$ in Fig.~\ref{fig1}(d), these peaks are positioned at $5.38t$ and $5.85t$. The energy difference between them is $0.47t\sim\omega_{0}$, and thus it is expected that the exciton has a polaronic feature. Because of the splitting, the line shape of the exciton is asymmetric (see Fig.~\ref{fig1}(d)). The shape is different from that obtained by a classical treatment of phonons~\cite{Iwano}. The energy gain of the exciton by $g$ is still smaller than $2g^{2}/\omega_{0}$, even though the exciton would be localized for $V=3t$.

Next, we examine whether the effect of $g$ on the exciton is enhanced or reduced by $U$. For this purpose, we compare the spectra for $U=10t$ and in the large-$U$ limit~\cite{note1}. It is usually recognized that the parameter $U=10t$ is characteristic of strong electron-electron correlation. Actually, for $g=0$, the absorption spectra for these parameters are almost the same. In this case, the exciton is bounded for $V\ltsim 2t$ and $U=10t$, and this critical $V$ value is close to the critical one $V=2t$ in the large-$U$ limit~\cite{Matsueda}. However, the line shape of the stress tensor correlation function changes dramatically even for $10t<U<40t$~\cite{Matsueda}. As will be shown later, we can see quantitative difference between the absorption spectra for $U=10t$ and in the large-$U$ limit in the presence of $g$. This means that the effect of $g$ on the spectra is sensitive to the value of $U$.

In order to study the optical spectra for $U\rightarrow\infty$, we introduce a holon-doublon model coupled with phonons. This model describes the charge sector of the photoexcited states in the large-$U$ limit of the Hubbard-Holstein model (\ref{H}). We drop the spin degrees of freedom because of the spin-charge separation in the ground state, and thus the numerical calculation becomes easier. The Hamiltonian is defined by
\begin{eqnarray}
H &=& -t\sum_{i}(h_{i}^{\dagger}h_{i+1}+d_{i}^{\dagger}d_{i+1}+{\rm H.c.}) + \frac{U}{2}\sum_{i}h_{i}^{\dagger}h_{i} \nonumber \\
&& + \frac{U}{2}\sum_{i}d_{i}^{\dagger}d_{i} - V\sum_{i}(h_{i+1}^{\dagger}h_{i+1}+h_{i-1}^{\dagger}h_{i-1})d_{i}^{\dagger}d_{i} \nonumber \\
&& + \omega_{0}\sum_{i}b_{i}^{\dagger}b_{i} + g\sum_{i}(b_{i}^{\dagger}+b_{i})(h_{i}^{\dagger}h_{i}-d_{i}^{\dagger}d_{i}),
\label{HD}
\end{eqnarray}
where $h_{i}^{\dagger}$ ($h_{i}$) and $d_{i}^{\dagger}$ ($d_{i}$) are creation (annihilation) operators of holon and doublon at site $i$, respectively. The holon and doublon obey the local constraint described by $h_{i}^{\dagger}h_{i}+d_{i}^{\dagger}d_{i}=1$. The current-current correlation function can be described by
\begin{eqnarray}
\chi(\omega)=-\frac{2}{\pi L}{\rm Im}\left(1\left|\frac{1}{\omega+E_{0}-H+i\gamma}\right| 1\right),
\end{eqnarray}
where $\left| 1\right)$ is a state with one holon-doublon pair. This state is created by applying the current operator $j=-it\sum_{i}(d_{i}^{\dagger}h_{i+1}^{\dagger}-h_{i}^{\dagger}d_{i+1}^{\dagger})$ to the classical Neel ordered state. Here, we introduce the prefactor $2$ in order to take account of the effect of quantum spin fluctuation.

\begin{figure}[htbp]
\begin{center}
\includegraphics[width=11cm]{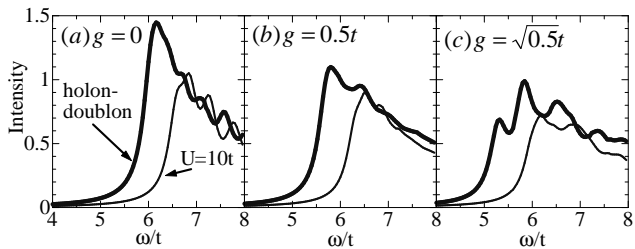}
\end{center}
\caption{Comparison of $\chi(\omega)$ between the Hubbard-Holstein model with $U=10t$ (fine solid line) and the holon-doublon model (bold solid line). We take $V=2t$. The spectra for the holon-doublon model are shifted for the comparison.}
\label{fig2}
\end{figure}

Figure~\ref{fig2} shows comparison of $\chi(\omega)$ between the Hubbard-Holstein model with $U=10t$ and the holon-doublon model. Here, $V$ is taken to be $2t$, and then the lowest-energy photoexcited state is an exciton. In order to examine the difference of the phonon effect on the exciton between these models, $g$ is changed from $0$ to $\sqrt{0.5}t$. For instance, the difference appears in Fig.~\ref{fig2}(c): the spectral weight of the lowest-energy peak is largest for $U=10t$, while the weight of the lowest-energy peak is smaller than that of the second-lowest-energy peak in the large-$U$ limit. Therefore, the increase of $U$ significantly changes the lineshape of the exciton in the presence of $g$. This means that the effecive EP interaction increases with $U$.

Let us consider why $U$ enhances the effect of the EP interaction on the exciton. First of all, it would be useful for later discussion to introduce the exactly solvable case of the holon-doublon model. In the large-$V$ limit, by the canonical transformation anologous to the method taken in ref~\cite{Mahan}, we have the following result
\begin{eqnarray}
\chi(\omega)=-\frac{1}{\pi L}{\rm Im}\sum_{l=0}^{\infty}\frac{A_{l}}{\omega-\Omega_{l}+i\gamma},
\end{eqnarray}
where $A_{l}=8\pi(2g^{2}/\omega_{0}^{2})^{l}\exp{(-2g^{2}/\omega_{0}^{2})}/l!$ and $\Omega_{l}=U-V-4t^{2}/V-2g^{2}/\omega_{0}+l\omega_{0}$. We find that the spectral weight for the lowest-energy photoexcited state $\left|n_{0}\right>$, $A_{0}=8\pi\exp{(-2g^{2}/\omega_{0}^{2})}$, decreases exponentially as $g$ increases. On the other hand, the decrease of the spectral weights for higher-energy photoexcited states $\left|n_{l}\right>$ is suppressed by a factor $(2g^{2}/\omega_{0}^{2})^{l}$. When $g>\omega_{0}/\sqrt{2}$, $A_{0}$ becomes smaller than $A_{1}$ because of strong renormalization of $\left|n_{0}\right>$ due to large EP coupling. In this case, we suppose large expectation value of the phonon number for $\left|n_{0}\right>$, $N_{0}=\left<n_{0}\right|\sum_{i}b_{i}^{\dagger}b_{i}\left|n_{0}\right>$. In the following consideration, we assume this parameter region where $\left|n_{0}\right>$ is strongly renormalized and $A_{0}<A_{1}$.

In the large-$U$ limit, charge fluctuation does not occur in the ground state, and $\left|n_{0}\right>$ is composed of the bases with one holon-doublon pair. Because of the inhomogeneous charge distribution, phonons are spontaneously excited around the pair, leading to large $N_{0}$. On the other hand, when $U$ is finite, the bases without the pair exist in $\left|n_{0}\right>$. Since these bases represent homonegeous charge distribution, $N_{0}$ decreases as $U$ decreases.

Next, we examine $A_{0}$ as a function of $N_{0}$ in order to understand the relation between $A_{0}$ and $U$. It is clear that the dominant contribution to the ground state comes from $\left| 0\right>\sim\left| AF\right>\left| 0\right)$, where $\left|AF\right>$ and $\left| 0\right)$ represent the antiferromagnetic state and phonon vacuum, respectively, as long as the ground state is the Mott insulator. In this case, $A_{0}=\left|\left<n_{0}\right|j\left|0\right>\right|^{2}$ is given by
\begin{eqnarray}
A_{0}\sim\left| t\sum_{i}\left<n_{0}\right|\left(\left| 0,\uparrow\downarrow\right>_{i,i+1}-\left|\uparrow\downarrow,0\right>_{i,i+1}\right)\left| 0\right)\right|^{2},
\end{eqnarray}
where the holon-doublon pair $\left|0,\uparrow\downarrow\right>$ is located on the $i$-th and $(i+1)$-th sites. This equation shows the relation between $A_{0}$ and $N_{0}$. When $N_{0}$ decreases, the weights of the bases without phonons become dominant in $\left|n_{0}\right>$, and this effect increases $A_{0}$. According to the previous paragraph where we found that $N_{0}$ decreases with decreasing $U$, $A_{0}$ increases with decreasing $U$. We have also confirmed $U$ dependence of $N_{0}$ and $A_{0}$ by exact diagonalization on small clusters, and the diagonalization result agrees well with the above consideration.

Therefore, it is concluded that the effect of the EP interaction on an exciton is very strong in correlated electron systems. This is essentially different from the effect of the EP interaction on the ground state where the Peierls instability is suppressed with the increase of the on-site Coulomb repulsion.

Here, we compare the present DMRG results with the optical absorption spectrum in the 1D cuprate Sr${}_{2}$CuO${}_{3}$~\cite{Ono}. The absorption spectrum has asymmetric structure: the high-energy side of the spectrum is broad, while the low-energy side is fitted with a single Lorentzian function with a broadening factor of the order of 0.1 eV. In addition, we find a flat part at the top of the absorption peak ($1.75 {\rm eV}\ltsim\omega\ltsim 1.82 {\rm eV}$). The energy range 0.07 eV of this flat part is comparable to a phonon frequency of the breathing mode seen in Ca${}_{1.8}$Sr${}_{0.2}$CuO${}_{3}$. We consider that the flat part is a signature of multiple phonon peaks, although the clear peak structures would be washed away by the phonon dispersion. In our previous DMRG calculation for the single-particle excitation spectra, the dimensionless coupling parameter is estimated to be $\lambda\sim0.25$ ($\omega_{0}=0.5t$ and $g=\sqrt{0.5}t$)~\cite{Matsueda3}. Figure~\ref{fig1}(c) or \ref{fig1}(d) with this parameter $\lambda$ is consistent with the absorption spectrum. It is noted that $V$ may be larger than that estimated by the model without $g$, since the EP interaction makes the spectrum broad. This fact might affect the nonlinear optical properties in the cuprate~\cite{Matsueda}.

In summary, we have examined the effect of the EP interaction on the exciton in the 1D Mott insulator. We have performed the DMRG calculation of the optical spectra in the 1D Hubbard-Holstein model at half filling. We have found that the effect of the EP interaction on the exciton is enhanced by increasing the on-site Coulomb repulsion. This enhancement is in contrast to the effect of the EP interaction on the ground state where the Peierls instability is suppressed by the on-site Coulomb repulsion. The DMRG result with $g$ is consistent with linear absorption spectrum of the 1D cuprate Sr${}_{2}$CuO${}_{3}$.

The authors thank G. Chifune for discussion. This work was supported by the Next Generation Supercomputing Project of Nanoscience Program, CREST, and Grant-in-Aid for Scientific Research from MEXT. A part of numerical calculations was performed in the supercomputing facilities in ISSP, University of Tokyo, YITP, Kyoto University, and IMR, Tohoku University. H. M. acknowledges hospitality of YKIS07 organized by the Yukawa International Program for Quark-Hadron Sciences at YITP.

\end{document}